# Electronic band structure, phonon spectrum, and elastic properties of LaOFeAs

M. Aftabuzzaman, A.K.M.A. Islam[1], S.H. Naqib

Department of Physics, Rajshahi University, Rajshah-6205, Bangladesh

**Abstract**

We present results of *ab-initio* calculations of the electronic band structure, lattice dynamical properties, and elastic constants of LaOFeAs, the parent compound of the recently discovered superconducting iron-oxypnictide ($LaO_{1-x}F_xFeAs$). The total and partial electronic density of states (EDOS) of the undoped LaOFeAs (in the insulating or metallic state) are extracted from the electronic band structure. The phonon dispersion and the phonon density of states (PDOS) are also studied. Possible implications of the band structure, EDOS, and PDOS of LaOFeAs on the eventual appearance of high-$T_c$ superconductivity upon carrier doping are discussed. The values of various independent elastic constants for both insulating and metallic states are estimated and discussed.

## 1. Introduction

Recently a family of rare-earth (RE) iron-oxypnictides (RE-OFeAs) of ZrCuSiAs structure has attracted great deal of interest from the superconductivity researchers. The discovery of superconductivity at $T_c$ = 26 K in the fluorine doped $LaO_{1-x}F_xFeAs$ [1] and subsequent attainment of a $T_c$ of 55 K in the related $SmO_{1-x}F_xFeAs$ compound [2] have generated a flurry of experimental and theoretical activities [3-8]. Like the superconducting cuprates, iron-oxypnictides are quasi two-dimensional (2D), parent materials are antiferromagnetic, and superconductivity occurs upon adding either electrons or holes in the FeAs layers. There is even a pseudogap in the quasiparticle excitation spectrum in these Fe based superconductors [9, 10]. Such similarities with the cuprates have initiated lively debates within the superconductivity community on whether a similar mechanism is at work in both the types [11, 12]. Common to many ideas is that Cooper pairing itself may be emergent from more than one competing phases driven by the presence of an underlying quantum critical point.

Understanding the ground state electronic and lattice dynamical properties of the parent compound is of paramount importance since the emergence of superconductivity on carrier doping or alternatively on suppression of the spin density wave (SDW) [13, 14] must be related to the presence of various interactions in the undoped strongly correlated electronic material. Elastic constants of iron-oxypnictides, on the other hand, are related to the mechanical properties and are of considerable importance in view of future technological applications of these materials.

It was found experimentally [15, 16] that the undoped compound shows metallic SDW state with a small ordered magnetic moment at ~ 135 K. Since the magnetic moment is highly reduced, a zero spin approximation has some justification, conductivity on the other hand shows 'bad metallicity' [17] and its magnitude is rather low for the undoped material. Therefore, several earlier first-principle calculations were done assuming LaOFeAs as an insulator [18-20]. Haule *et al.* [21] computed the electronic structure of $LaO_{1-x}F_xFeAs$ within the combination of the Density Functional Theory (DFT) and the Dynamical Mean Field Theory (DMFT). They have found that the compound in the normal state is a strongly correlated metal and the parent compound is a bad metal at the verge of a metal insulator transition.

---

[1] Corresponding author.
*E-mail address*: azi46@ru.ac.bd (A.K.M.A. Islam).

In this study calculations are made assuming undoped LaOFeAs as an insulator as well as a metal. We have performed first-principle density functional calculations of the electronic band structure, EDOS, phonon dispersion, PDOS, and the various independent elastic constants for the undoped LaOFeAs. We have analyzed the electronic band structure and the phonon spectrum and discussed the possible implications of these on the occurrence of superconductivity in the doped compounds.

## 2. Method of computations

The *ab-initio* calculations were performed using the CASTEP program [22]. The geometrical optimization was done for LaOFeAs (space group 129 *P*4/*nmm*) treating the system as non-spin polarized in both insulating and metallic states. The generalized gradient approximation (GGA) of Perdew, Burke, and Ernzerhof (PBE) [23] potentials have been incorporated for the simulation. We have used a 10×10×4 Monkhorst grid to sample the Brillouin zone. All structures have been fully optimized until internal stress and forces on each atom are negligible. For all relevant calculations the plane wave basis set cut-off used is 380 eV and the convergence criterion is $0.5 \times 10^{-5}$ eV/atom.

Calculations of phonon spectra, electron-phonon (e-ph) coupling and phonon density of states were performed using plane waves and pseudopotentials with QUANTUM ESPRESSO [24]. We employed ultrasoft Vanderbilt pseudopotentials [25], with a cut-off of 50 Ryd for the wave functions, and 400 Ryd for the charge densities. The *k*-space integration for the electrons was approximated by a summation over a 8×8×4 uniform grid in reciprocal space, with a Gaussian smearing of 0.01 Ryd for self-consistent cycles; a much finer (16×16×8) grid was used for evaluating the phonon DOS and the e-ph linewidths. Dynamical matrices and e-ph linewidths were calculated on a uniform 4×4×2 grid in phonon *q*-space. Phonon dispersions and DOS were then obtained by Fourier interpolation of the dynamical matrices, and the Eliashberg function by summing over individual linewidths and phonons.

## 3. Results and discussion

### 3.1. Geometrical optimization

LaOFeAs has a layered tetragonal crystal structure as shown in Fig. 1. Layers of La-O are sandwiched between layers of Fe-As. The Fe atoms form a square two dimensional lattice with Fe-Fe lattice spacing of 2.85 Å (assuming a metallic state). The obtained lattice parameters and internal atomic positions of LaOFeAs are given in Table 1 using non-spin polarized calculation. These data are in very good agreement with the experimental results due to Kamihara *et al.* [1] and those determined in Rietveld analysis for undoped LaOFeAs (*P*4/*nmm*) by Nomura *et al.* [26]. Fe and O atoms sit at the centre of slightly distorted As and La tetrahedra; the As tetrahedra are squeezed in the *z* direction, so that there are two As-Fe-As angles ($\theta_1$, $\theta_2$), which are either larger or smaller than the regular tetrahedron value ($\theta = 109.47$ deg).

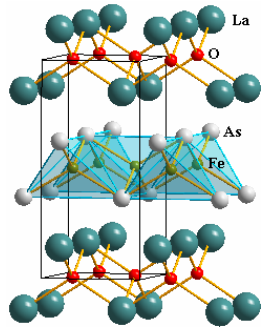

Fig. 1. Tetragonal crystal structure of LaOFeAs compound. FeAs layers are formed by Fe ions tetrahedraly surrounded by As ions.

Table 1
Structural parameters of LaOFeAs.

| | $a$ (Å) | $c$ (Å) | $V$ (Å$^3$) | $z_{La}$ | $z_{As}$ | $\theta_1$ (deg) | $\theta_2$ (deg) |
|---|---|---|---|---|---|---|---|
| This (insul.) | 4.0277 | 8.6699 | 140.65 | 0.1490 | 0.6357 | 104.7 | 119.4 |
| This (metal) | 4.0310 | 8.6490 | 140.54 | 0.1489 | 0.6353 | 104.6 | 119.7 |
| Theo. [27] | 3.996 | 8.636 | 137.9* | 0.1413 | 0.6415 | 105.8 | 117.1 |
| Expt. [1] | 4.035 | 8.741 | 142.31* | 0.1415 | 0.6512 | 107.5 | 113.5 |
| Expt. [26] | 4.03268 | 8.7411 | 142.15 | 0.1415 | 0.6512 | | |

*calculated value.

## 3.2. Electronic band structure and DOS

The material properties can be understood if one can identify the character of dominant bands near the Fermi level, their energy etc. The calculated bands along high symmetry directions of the Brillouin zone and the DOS for pure insulating and metallic states of LaOFeAs are shown in Fig. 2. The bands near the Fermi level distinguish between the two states of the compound.

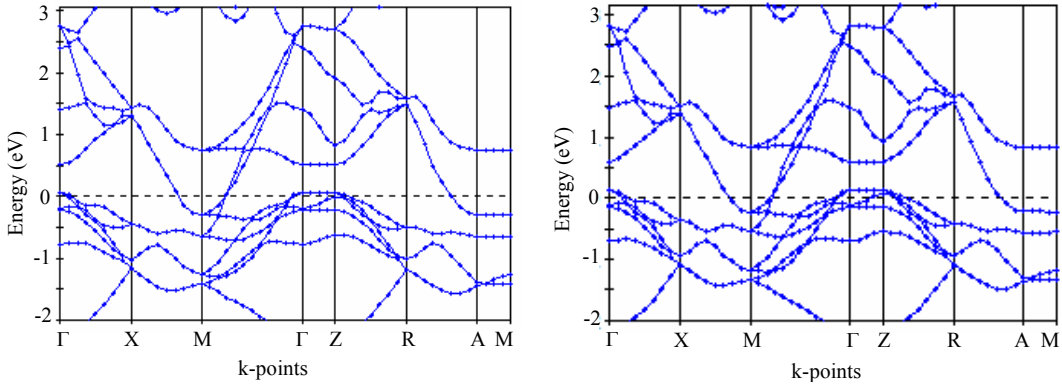

Fig. 2. Band structure of undoped LaOFeAs assuming insulator (*left*), metal (*right*).

We observe from Fig. 2 that a small dispersion along the c-axis (from Γ to Z and from A to M) which is indicative of weak interactions between LaO and FeAs layers and a 2-dimensional nature of the compound. The bands below $E_F$ are relatively flat and have little contribution coming from the As-*p* states, while the bands above $E_F$ are quite dispersive, and have large weight coming from the As-*p* character (as discussed below).

The total and projected density of states of LaOFeAs are shown in Fig. 3 in order to discuss the electronic structure further. It is seen that the dominant states at the Fermi level come mainly from Fe 3*d* atomic states extending roughly between -2.2 eV and 2 eV. This band width of ~ 4.2 eV is ~ 60% that of metallic bcc Fe, which may suggest significant itinerant character of *d* states [28]. The state energy for O-*p* and As-*p* orbitals are below that for Fe-3*d* and lie within ~ -5.5 eV to -2.2 eV. The Fermi level of the system in the metallic state cuts the band structure in a region where the DOS is 4.0 (2.08, insulating state) states/eV/unit cell. This value is high but decreases rather rapidly. Boeri *et al.* [27] found an energy opening in the electronic spectrum around 0.2 eV. It was indicated in earlier works [4, 21] that such a high DOS at the Fermi level drives the system close to a magnetic instability.

We note that electronic states of La-O layers are rather far from the Fermi level. Nekrasov *et al.* [29] have also shown that the electronic structure of RE-OFeAs compounds (e.g. the DOS around the Fermi level) practically does not depend on the kind of rare-earth ion used in a wide energy interval around the Fermi level, which is relevant for superconductivity in FeAs layers. Thus the *p-d* hybridization between O and Fe (as illustrated by partial DOS contributions shown in Fig. 4) is negligible. But within the FeAs layers a quite sizeable mixing between Fe-*d* and As-*p* is apparent around -3.2 eV where As 4*p* band is strongly peaked (see Fig. 4). Therefore, LaOFeAs can be considered as a quasi-2D which looks to be similar to the case of conventional high-$T_c$ cuprates.

It may be pointed out here that the simple tetrahedral crystal field predicts that the five *d* orbitals split into low-lying two-folds $e_g$ ($d_{x^2-y^2}$ and $d_{3z^2-1}$) states and up-lying three-folds $t_{2g}$ states. The distortion of the tetrahedron from its normal shape (see Table 1) will further split the $e_g$ and $t_{2g}$ manifolds significantly making the final orbital distributions complicated [27, 30]. The manifold is two-folds, between which an energy gap of about 0.5 eV exists. The nominal number of *d* electrons in LaOFeAs is 6, and the low-lying three-fold bands are nearly fully occupied with the Fermi level, $E_F$, located close to the energy gap, and therefore the nonmagnetic metallic ground state is stabilized [30].

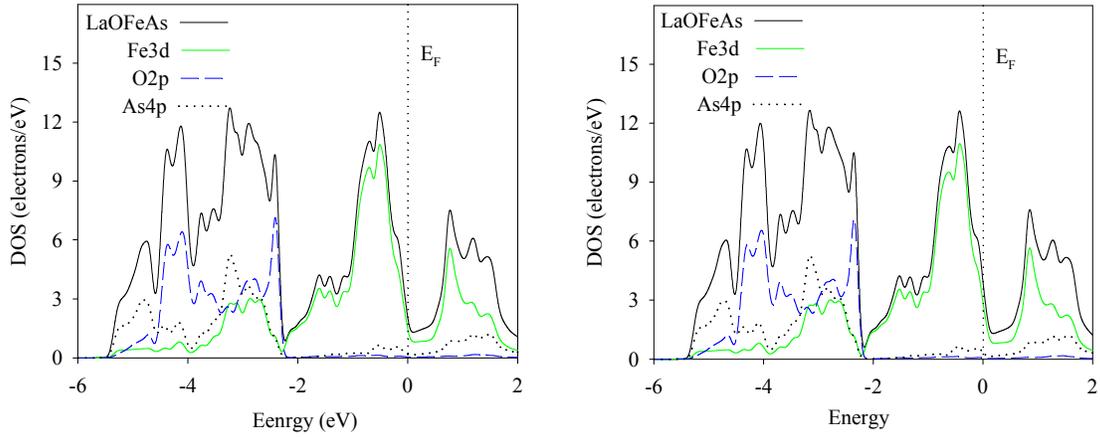

Fig. 3. Density of states (DOS) of undoped LaOFeAs as insulator (*left*), metal (*right*).

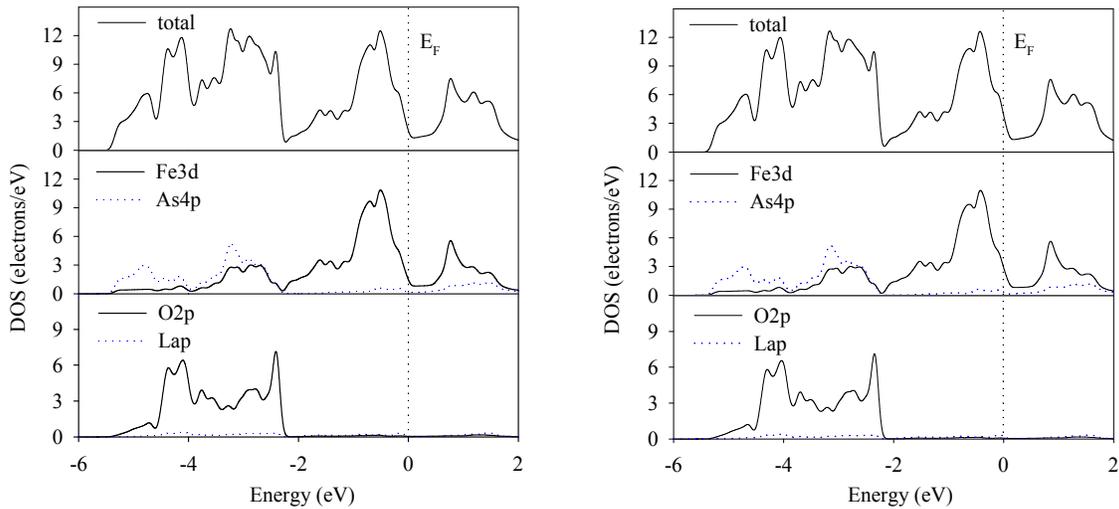

Fig. 4. Total (*upper panel*) and partial densities of states for LaOFeAs as insulator (*left*), metal (*right*).

*3.3. Phonon spectrum and electron-phonon coupling*

Fig. 5 shows the calculated phonon dispersions and the corresponding phonon density of states of LaOFeAs in the metallic state. The phonon dispersions show a set of 24 phonon branches since there are two formula units per primitive cell. The spectrum extends up to 500 cm$^{-1}$; it shows three main peaks centred at ~ 105 cm$^{-1}$ and smaller peaks at ~ 180 cm$^{-1}$ and ~ 265 cm$^{-1}$. Similar observations were reported earlier by other workers [4, 27]. The higher frequency manifold is due mainly from vibrations of O atoms which are dispersive and well separated in energy from those other atomic species [27]. The lower manifold from 0 to 200 cm$^{-1}$ consists of the acoustic modes and modes of mixed, but mainly metallic character. Further the projection on in- and out-of-plane modes (not shown) does not show any clear separation between the patterns of vibration. This is contrary to what is expected in a layered compound like LaOFeAs. Thus the three major peaks cannot be attributed to a single vibration pattern [27].

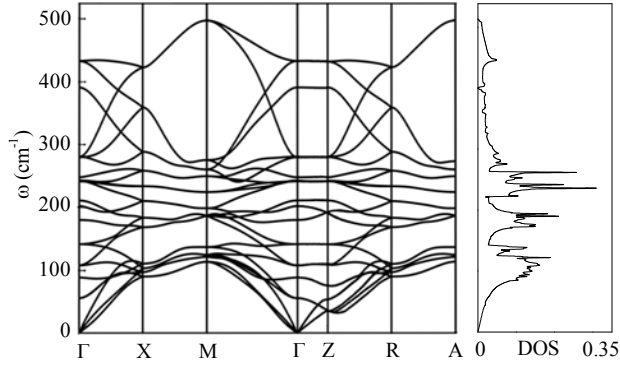

Fig. 5. Electron-phonon properties of LaOFeAs. *Left:* Phonon dispersion relations; *Right:* Phonon DOS.

The frequency-dependent e-ph coupling constant $\lambda(\omega)$ and Eliashberg function $\alpha^2F(\omega)$ are given by [31]

$$\lambda(\omega) = 2\int_0^\omega d\Omega\, \alpha^2 F(\Omega)/\Omega \qquad (1)$$

$$\alpha^2 F(\omega) = \frac{1}{N(0)} \sum_{nmk} \delta(\varepsilon_{nk}) \delta(\varepsilon_{mk+q}) \sum_{vq} |g_{v,nk,m(k+q)}|^2 \delta(\omega - \omega_{vq}) \qquad (2)$$

where $g_{v,nk,mk'}$ are the bare DFT matrix elements.

The total e-ph coupling constant $\lambda$ for metallic state of LaOAsFe, is obtained by numerical integration of Eq. (1) up to $\omega = \infty$ and is shown Table 2.

Table 2
Electron-phonon properties.

|      | $N(0)$ eV$^{-1}$f.u$^{-1}$ | $\omega_{ln}$ (K) | $\lambda$ | $\mu^*$ | $T_c$ (K) |
|------|---------|---------|---------|---------|---------|
| This | 2.16    | 221     | 0.13    | 0       | 0.12    |
|      |         |         |         | 0.12    | 0       |
| [27] | 2.1     | 205     | 0.21    | 0       | 0.5     |

The resulting value of $\lambda$ is almost the same as that obtained earlier by Boeri *et al.* [27] and Mazin *et al.* [32]. The logarithmically-averaged frequency $\omega_{ln}$, is calculated to be 221 K. We use $\mu^* = 0$ and 0.12 in the simplified Allen-Dynes formula for $T_c$ [33]:

$$k_B T_c = \frac{\hbar \omega_{ln}}{1.2} \exp\left[-\frac{1.04(1+\lambda)}{\lambda - \mu^*(1+0.62\lambda)}\right] \qquad (3)$$

This gives $T_c = 0.12$ K for $\mu^* = 0$ (Table 2) which is too small to explain the $T_c = 26$ K observed in doped sample. Boeri *et al.* [27] noted that the undoped LaOFeAs is close to a magnetic instability, due to the presence of a very sharp peak in the electronic density of states. The system moves away from the magnetic instability on doping with electrons. This causes the reduction of the DOS at the Fermi level, without changing the band structure appreciably and without introducing new bands at $E_F$. Thus the net result of such doping would be to reduce the value of $\lambda$. Therefore, the value $\lambda = 0.13$ for the undoped material is actually an upper bound for the value in the e-doped compound. This value is lower than what is seen in any known e-ph superconductor, e.g., $\lambda = 0.44$ in metallic aluminium ($T_c = 1.3$ K).

*3.4. Elastic constants*

Table 3 shows the calculated values of the six independent elastic constants, $C_{ij}$ for the tetragonal LaOFeAs along with the results obtained by Shein and Ivanovskii [19]. The elastic constants are all positive and satisfy the well-known Born's criteria for tetragonal crystals: $C_{11} > 0$, $C_{33} > 0$, $C_{44} > 0$, $C_{66} > 0$, $(C_{11} - C_{12}) > 0$, $(C_{11} + C_{33} - 2C_{13}) > 0$ and $[2(C_{11} + C_{12}) + C_{33} + 4C_{13}] > 0$. All these imply the mechanical stability of the crystal.

LaFeAsO is usually prepared and investigated as polycrystalline ceramics in the form of aggregated mixtures of micro-crystallites with a random orientation. Thus it is quite useful to estimate the corresponding parameters for these polycrystalline materials. The theoretical polycrystalline elastic moduli for LaOFeAs may be calculated from the set of six independent elastic constants. Hill [34] proved that the Voigt and Reuss equations represent upper and lower limits of the true polycrystalline constants. He showed that the polycrystalline moduli are the arithmetic mean values of the moduli in the Voigt ($B_R$, $G_R$) and Reuss ($B_R$, $V_R$) approximation, and are thus given by $B_H \equiv B = \frac{1}{2}(B_R + B_V)$ and $G_H \equiv G = \frac{1}{2}(G_R + G_V)$. The expression for Reuss and Voigt moduli can be found elsewhere [35, 36]. From the calculated values of the moduli it is easy to evaluate the compressibility ($\beta$), Young's moduly (Y), and the Poisson's ratio ($v$) (see Table 4).

Table 3
Calculated elastic constants, $C_{ij}$ for monocrystalline LaOFeAs phase. Bulk and shear moduli ($B_V$, $G_V$) are for macroscopic material in the Voigt approximation.

| $C_{ij}$ GPa | LaOFeAs | | |
|---|---|---|---|
| | This | | [19] |
| | Insulator | Metal | Insulator |
| $C_{11}$ | 218.5 | 213.6 | 191.9 |
| $C_{12}$ | 75.5 | 80.4 | 55.9 |
| $C_{13}$ | 87.8 | 93.7 | 61.6 |
| $C_{33}$ | 151.0 | 138.2 | 144.8 |
| $C_{44}$ | 63.8 | 61.9 | 44.1 |
| $C_{66}$ | 68.0 | 69.0 | 77.9 |
| $B_v$ | 121.1 | 122.3 | 98.5 |
| $G_v$ | 61.6 | 58.4 | 56.5 |

Table 4
Calculated elastic parameters for polycrystalline LaOFeAs.

|  | Bulk moduli $B$ (GPa) | Compressibility $\beta$ (GPa$^{-1}$) | Shear moduli $G$ (GPa) | Young's moduli $Y$ (GPa) | Poisson ratio $v$ |
|---|---|---|---|---|---|
| Insulator (This) | 119.7 | 0.00835 | 60.2 | 154.7 | 0.285 |
| Metal (This) | 120.2 | 0.00832 | 55.9 | 145.2 | 0.298 |
| Insulator [19] | 97.9 | 0.01022 | 56.2 | 141.5 | 0.259 |

The results obtained are presented in Table 3 from which we see that $B_v > G_v$. The shear modulus $G_v$ is parameter which limits the mechanical stability of a material. This is apparent due to the layered structure of the compound.

As compared with other superconductors, LaOFeAs is a soft material. This is evident from the value of relatively smaller bulk moduli for LaOFeAs (~120 GPa), which is smaller than the bulk moduli of other well known superconducting materials such as $MgB_2$, $MgCNi_3$, YBCO and $YNi_2B_2C$ [37-40].

Finally, the criterion of brittleness of a material is $B/G < 1.75$ [41]. According to this LaOFeAs lies just above the border of brittleness as $B/G$ is 2.15 (metal) and 1.99 (insulator). These values can be compared with 1.74 (insulator) obtained by Shein and Ivanovskii [19]. It is known that the values of the Poisson Ratio ($v$), minimal for covalent materials ($v = 0.1$), increase for ionic systems [42]. In our case, the value of $v$ for LaOFeAs assuming metal is 0.298, which is indicative of sizable ionic contribution in intra-atomic bonding.

**4. Conclusion**

The lattice parameters and internal atomic positions of LaOFeAs using non-spin polarized calculation are in very good agreement with the experimental results. The electronic band structure reveals that the dominant states at the Fermi level come mainly from Fe 3$d$ atomic states extending roughly between -2.2 eV and 2 eV. The Fermi level is located at the shoulder just below the dip in the $d$ band. We find a high density of states which decreases rapidly. Earlier works indicated that such a high DOS at the Fermi level drives the system close to a magnetic instability.

The electronic states of La-O layers are rather far from the Fermi level. The $p$-$d$ hybridization between the lighter most O and Fe is negligible. But within the FeAs layers a quite sizeable mixing between Fe-$d$ and As-$p$ is apparent around -3.2 eV where As 4$p$ band is strongly peaked. The strongly delocalized character of the Fe-$d$ states at ±2 eV around the Fermi level is responsible for very small e-ph matrix elements yielding a small $\lambda$. We see that such behaviour is an intrinsic property of this LaOFeAs, which is unlikely to be changed by external parameters, such as pressure or doping [27]. To explain the observed $T_c$ in the system one would need further mechanism beyond the conventional e-ph theory.

The analysis of elastic constants shows that LaOFeAs is a mechanically stable anisotropic material. The phase is a soft material (high compressibility) and lies just above the borderline of brittleness. Sizable ionic contribution in intra-atomic bonding in the compound is also predicted.

**Acknowledgements**

The authors would like to thank Prof. Eyvaz Isaev, Department of Theoretical Physics, Moscow Institute of Steel and Alloys, Russia, for useful help during the work.

## References

[1] Y. Kamihara, T. Watanabe, M. Hirano, and H. Hosono, J. Am. Chem. Soc. 130 (2008) 3296. doi:10.1021/ja800073m
[2] J. Yang, Z. -C. Li, W. Lu, W. Yi, X. -L. Shen, Z.-A. Ren, G. -C. Che, X.-L. Dong, L. -L. Sun, and F. Zhou, Supercond. Sci. Technol. 21 (2008) 082001. doi:10.1088/0953-2048/21/8/082001
[3] A. B. Vorontsov, M. G. Vavilov, and A. V. Chubukov, Phys. Rev. B 79 (2009) 060508(R). doi:10.1103/PhysRevB.79.060508
[4] D. J. Singh and M. -H. Du, Phys. Rev. Lett. 100 (2008) 237003. doi:10.1103/PhysRevLett.100.237003
[5] L. S. Mazov, arXiv: 0805.4097.
[6] A. V. Chubukov, D. Efremov, and I. Eremin, Phys. Rev. B 78 (2008) 134512. doi:10.1103/PhysRevB.78.134512
[7] M. D. Johanness and I. I. Mazin, arXiv: 0904.3857.
[8] P. V. Sushko, A. L. Schluger, M. Hirano, and H. Hosono, Phys. Rev. B 78 (2008) 172508. doi:10.1103/PhysRevB.78.172508
[9] Y. Ishida, T. Shimojima, K. Ishizaka, T. Kiss, M. Okawa, T. Togashi, S. Watanabe, X.-Y. Wang, C.-T. Chen, Y. Kamihara, M. Hirano, H. Hosono, and S. Shin, arXiv: 0805.2647v1.
[10] K. Ishida, Y. Nakai, and H. Hosono, arXiv: 0906.2045v1.
[11] I. I. Mazin and M. D. Johanness, arXiv: 0807.3737.
[12] E. Berg, S. A. Kivelson, and D. J. Scalapino, arXiv: 0905.1096v1.
[13] Z. Zhong, Q. Zhang, P. X. Xu, and P. J. Kelly, arXiv: 0810.3246v1.
[14] D. J. Singh, arXiv: 0901.2149v1.
[15] R. H. Liu, G. Wu, T. Wu, D. F. Fang, H. Chen, S. Y. Li, K. Liu, Y. L. Xie, X. F. Wang, R. L. Yang, L. Ding, C. He, D. L. Feng, and X. H. Chen, Phys. Rev. Lett. 101 (2008) 087001. doi:10.1103/PhysRevLett.101.087001
[16] T. M. McQueen, M. Regulacio, A. J. Williams, Q. Huang, J. W. Lynn, Y. S. Hor, D. V. West, M. A. Green, and R. J. Cava, Phys. Rev. B 78 (2008) 024521. doi:10.1103/PhysRevB.78.024521
[17] C. Wang, S. Jiang, Q. Tao, Z. Ren, Y. Li, L. Li, C. Feng, J. Dai, G. Cao and Z. Xu, arXiv: 0811.3925v1.
[18] I. R. Shein and A. L. Ivanovskii, arXiv: 0807.3422.
[19] I. Shein, A. Ivanovskii, Scripta Materialia 59 (2008) 1099. doi:10.1016/j.scriptamat.2008.07.028
[20] I. R. Shein and A. L. Ivanovskii, arXiv: 0807.0984.
[21] K. Haule, J. H. Shim, and G. Kotliar, arXiv:0803.1279 v1.
[22] S. J. Clark, M. D. Segall, C. J. Pickard, P. J. Hasnip, M. J. Probert, K. Refson, and M. C. Payne, Zeitschrift fuer Kristallographie 220 (2005) 567. doi:10.1524/zkri.220.5.567.65075
[23] J. P. Perdew, S. Burke, and M. Ernzerhof, Phys. Rev. Lett. 77 (1996) 3865. doi:10.1103/PhysRevLett.77.3865
[24] P. Giannozzi *et al*., http://www.quantum-espresso.org.
[25] D. Vanderbilt, Phys. Rev. B 41 (1990) 7892(R). doi:10.1103/PhysRevB.41.7892
[26] T. Nomura, S. W. Kim, Y. Kamihara, M. Hirano, P. V. Sushko, K. Kato, M.Takata, A. L. Shluger, and H.Hosono, cond-mat/arXiv:0804.3569; and also J. Phys. Soc. Jpn. Suppl. C 77 (2008) 32.
[27] L. Boeri, O. V. Dolgov, and A. A. Golubov, Phys. Rev. Lett. 101 (2008) 026403; L. Boeri, O. V. Dolgov, A. A. Golubov, arXiv:0902.0288v1.
[28] S. Ishibashi, K. Terakura, and H. Hosono, J. Phys. Soc. Jpn. 77 (2008) 053709.

doi:10.1143/JPSJ.77.053709
[29] I. A. Nekrasov, Z. V. Pchelkina, M. V. Sadovskii, Pis'ma v ZhETF, 87 (2008) 647.
[30] G. Xu, W. Ming, Y. Yao, X. Dai, S -C. Zhang, and Z. Fang, arXiv:0803.1282v2.
[31] G. M. Eliashberg, Sov. Phys. JETP 11, 696 (1960) and G. M. Eliashberg, Zh. Eksp. Teor. Fiz. 38 (1960) 966.
[32] I. I. Mazin, D. J. Singh, M. D. Johannes, and M. H. Du, arXiv:0803.2740 (2008).
[33] P. B. Allen and R. C. Dynes, Phys. Rev. B 12 (1975) 905. doi:10.1103/PhysRevB.12.905
[34] R. Hill, Proc. Phys. Soc. London A 65 (1952) 349. doi:10.1088/0370-1298/65/5/307
[35] A. Reuss, Z. Angew. Math. Mech. 9 (1929) 49. doi:10.1002/zamm.19290090104
[36] W. Voigt. Lehrbuch der Kristallphysik, Teubner, Leipzig. 1928.
[37] S. Meenakshi, V. Vijayakumar, R.S. Rao, B.K. Godwal, S.K. Sikka, P. Ravindran, Z. Hossian, R. Nagarajan, L.C. Gupta, R. Vijayaraghavan, Phys.Rev. B 58 (1998) 3377. doi:10.1103/PhysRevB.58.3377
[38] U. Jaenicke-Rossler, G. Zahn, P. Paufler, B. Wehner, H. Bitterlich, Physica C 305 (1998) 209. doi:10.1016/S0921-4534(98)00327-X
[39] H. Y. Wang, X. R. Chen, W. J. Zhu, Y. Cheng, Phys. Rev. B 72 (2005) 172502. doi:10.1103/PhysRevB.72.172502
[40] I. R. Shein, V. V. Bannikov, A. L. Ivanovskii, Physica C 468 (2008) 1. doi:10.1016/j.physc.2007.08.004
[41] S. F. Pugh, Phil. Mag. 45 (1954) 833.
[42] J. Haines, J. M. Leger, G. Bocquillon, Ann. Rev. Mater. Res. 31(2001) 1. doi:10.1146/annurev.matsci.31.1.1